\documentclass[longtitle,preprint,5p,times,twocolumn]{elsarticle}

\usepackage{amssymb}

\usepackage{caption}
\usepackage{xspace}
\usepackage{longtable}
\usepackage{pdflscape}
\usepackage{multirow}
\usepackage{ragged2e}
\usepackage{booktabs}
\usepackage{hyperref}
\usepackage{tikz}
\usetikzlibrary{arrows,chains,shapes,matrix,scopes,positioning,fadings,backgrounds,fit,mindmap,trees,decorations.markings,decorations.pathreplacing,calc}
\usetikzlibrary{decorations.pathmorphing,decorations.markings,trees,arrows.meta}

\usepackage{lineno}

\journal{Nuclear Instruments and Methods A}

\begin{document}

\begin{frontmatter}
\title{Scientific Computing Plan for the ECCE Detector at the Electron Ion Collider}
\def\theaffn{\arabic{affn}} 
\author[CFNS,StonyBrook,RBRC]{J.~C.~Bernauer}
\author[LANL]{C.~T.~Dean}
\author[MIT]{C.~Fanelli}
\author[BNL]{J.~Huang}
\author[BNL]{K. Kauder}
\author[JLab]{D.~Lawrence}
\author[ORNL]{J.D.~Osborn}
\author[MIT]{C.~Paus}

\author[MoreheadState]{J.~K.~Adkins}
\author[RIKEN]{Y.~Akiba}
% ...

%
%

%
%
%
%
%

% ...
\affiliation[BNL]{organization={Brookhaven National Laboratory},
	 city={Upton},
	 postcode={11973},
	 state={NY},
	 country={USA}} 
 
% ...
\affiliation[JLab]{organization={Thomas Jefferson National Accelerator Facility},
	 addressline={12000 Jefferson Ave.},
	 city={Newport News},
	 postcode={24450},
	 state={VA},
	 country={USA}} 
 
% ...
\affiliation[LANL]{organization={Los Alamos National Laboratory},
	 city={},
	 postcode={},
	 state={NM},
	 country={USA}} 
 
% ...
\affiliation[MoreheadState]{organization={Morehead State University},
	 city={Morehead},
	 postcode={},
	 state={KY},
	 }
 
\affiliation[MIT]{organization={Massachusetts Institute of Technology},
	 addressline={77 Massachusetts Ave.},
	 city={Cambridge},
	 postcode={02139},
	 state={MA},
	 country={USA}} 
 
% ...
\affiliation[ORNL]{organization={Oak Ridge National Laboratory},
	 addressline={PO Box 2008},
	 city={Oak Ridge},
	 postcode={37831},
	 state={TN},
	 country={USA}} 
 
% ...
\affiliation[RIKEN]{organization={RIKEN Nishina Center},
	 city={Wako},
	 state={Saitama},
	 country={Japan}} 
 
% ...
\affiliation[CFNS]{organization={Center for Frontiers in Nuclear Science},
	 city={Stony Brook},
	 postcode={11794},
	 state={NY},
	 country={USA}} 
 
\affiliation[StonyBrook]{organization={Stony Brook University},
	 addressline={100 Nicolls Rd.},
	 city={Stony Brook},
	 postcode={11794},
	 state={NY},
	 country={USA}} 
 
\affiliation[RBRC]{organization={RIKEN BNL Research Center},
	 city={Upton},
	 postcode={11973},
	 state={NY},
	 country={USA}} 
% ...

\begin{abstract}

The Electron Ion Collider (EIC) is the next generation of precision QCD facility to be built at Brookhaven National Laboratory in conjunction with Thomas Jefferson National Laboratory. There are a significant number of software and computing challenges that need to be overcome at the EIC. During the EIC detector proposal development period, the ECCE consortium began identifying and addressing these challenges in the process of producing a complete detector proposal based upon detailed detector and physics simulations. In this document, the software and computing efforts to produce this proposal are discussed; furthermore, the computing and software model and resources required for the future of ECCE are described. 

\end{abstract}
\end{frontmatter}

\setcounter{tocdepth}{1}
\tableofcontents

\section {Introduction}
\label{sec:introduction}


The Electron Ion Collider (EIC) is the next generation precision nuclear physics accelerator complex that is being constructed in the United States. The EIC is expected to start producing data in the early 2030's, and is unique as it will collide high energy polarized electrons with polarized protons and a wide range of nuclei. As such, it will introduce new paradigms in large scale nuclear physics experiments. Expected luminosities at the EIC will reach upwards of $10^{34}$ cm$^{-2}$s$^{-1}$; consequently, there will be an extremely large data sample to process. Recent efforts from modern collider physics experiments have shown the benefits of near real-time analysis~\cite{Benson_2015,Aaij_2019}. Therefore, there is a strong desire to develop software and computing infrastructure that reliably and quickly processes data for analysis.

As the EIC will start taking data in nearly a decade, there are a number of new paradigms that have the opportunity to be explored in this software R\&D phase. An example of such a new paradigm is the use of Artificial Intelligence (AI) and machine learning (ML). The EIC has the unique opportunity to be one of the first large-scale facilities to systematically incorporate and employ AI, starting from the detector design and R\&D phases. The relevance of AI for the EIC has been highlighted among the \textit{Opportunities for Computing} of the EIC Yellow Report \cite{YellowReport}. AI potentially permeates all aspects of this Computing Plan; in fact, it already plays a significant role in the design and R\&D phases of the EIC~\cite{cisbani2020ai}~\cite{AI4EIC2021}.

This document presents a proposed computing plan for the EIC Comprehensive Chromodynamics Experiment (ECCE) detector at the EIC~\cite{YellowReport}. This includes estimates of the rates from the detector, the pipeline for processing and storing the data, and how the collaboration members will access the data. Software systems for monitoring, calibration, reconstruction, and analysis are discussed. Estimates of the computing and storage requirements are included. AI detector optimization techniques are also discussed as this is expected to be a large part of the computing effort over the next few years. While we attempt to include some forward thinking plans in this regard, we necessarily do need to rely on past experience with other large experiments such as sPHENIX~\cite{sphenix_computing_plan_2019} and LHCb~\cite{CAMPORAPEREZ2016280} to serve as guides.

\section {Online}
\label{sec:online}

\subsection{Data acquisition}

\begin{figure*}[th]
 \begin{center}
  \input{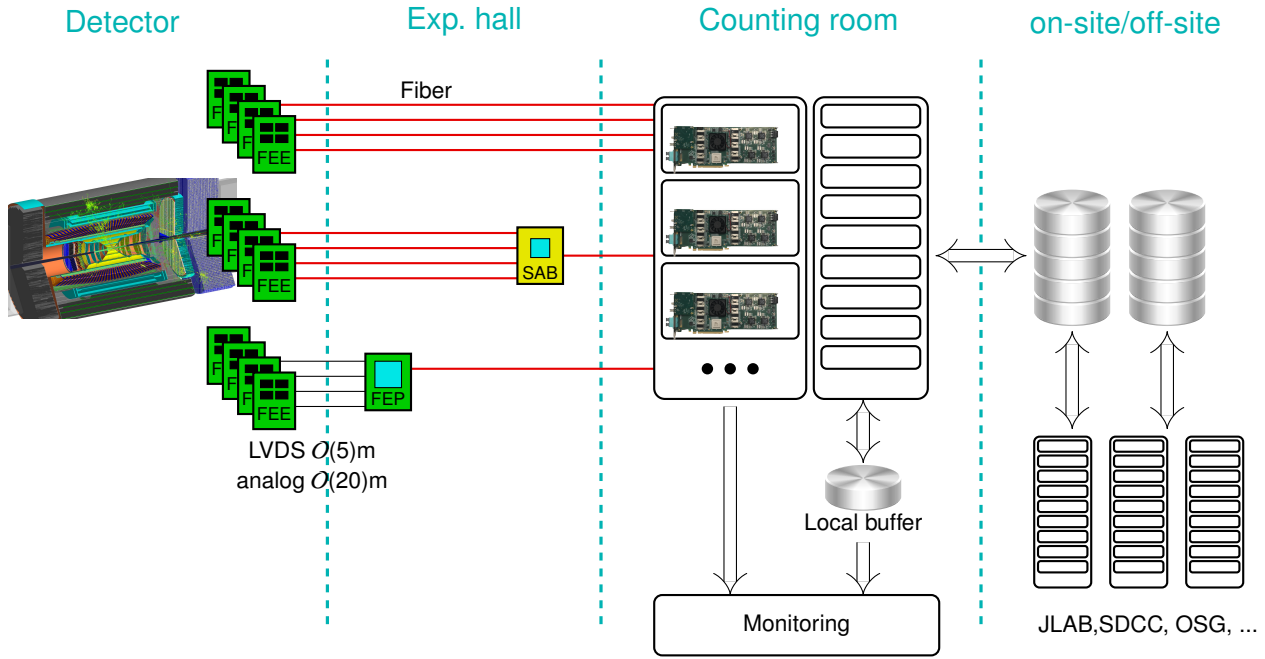}
  
  \caption[Data Acquisition Diagram]{\label{fig:data_acquisition_diagram} The DAQ electronics deployment can be roughly divided by their location, with Front End Electronics (FEE) modules on/near the detector; Front End Processor boards (FEP) which digitize or reformat detector information and Stream Aggregator Boards (SAB), which bundle streams, in the hall; and online filtering and monitoring in the counting room. Long term storage and analysis processing is performed in a federated model on multiple sites. }
 \end{center}
\end{figure*}

We envision a DAQ system following a streaming readout paradigm, where all data is collected in an unbiased hardware trigger-less system. In the following, we will describe the individual components as well as the overall data flow and bandwidth model. An overview of the system is shown in Figure \ref{fig:data_acquisition_diagram}.

\subsection{DAQ components}
 Front-end electronics (FEE) modules sit inside or on the detector. In most cases, detector-specific ASICs provide the data conversion from the analog to digital domain, do zero-suppression and provide an interface to fiber transceivers for data transport to the counting room. For this, we envision Front-End Link eXchange (FELIX)-type PCIe-based receiver cards\footnote{In the following, we will use ``FELIX card'' as a stand-in for a successor board of similar architecture.} which support a large number of high speed fiber links per card~\cite{Chen:2019owc}. FELIX is designed for new or upgraded detectors and trigger systems in the ATLAS Phase-I upgrade and High-Luminosity LHC, and is implemented by server PCs with commodity network interfaces and PCIe cards with large field-programmable gate arrays (FPGAs) and many high speed serial fiber transceivers. 

Since some FEE may not utilize the full bandwidth of a fiber link, cost-effective stream aggregator boards (SABs), based either on small FPGAs or COTS multiplexer ICs, can bundle multiple fiber links coming from FEE to a single fiber connected to the FELIX cards. 

Because of space and services constraints, or because no suitable ASIC is available, some FEEs will connect to Front end processor (FEP) modules via digital (LVDS) or analog links. The FPGA and possibly analog to digital converter logic on the FEP will then generate a data stream suitable for fiber transport to the FELIX cards.

In the counting room, we expect a number of special servers which house the FELIX cards. Each FELIX/CPU combination sees data from a certain subset of detector channels and can do additional data reduction before sending out the data to the counting room CPU farm. This CPU farm (with possibly GPU accelerators) will do further data reduction, for example via high level data selection algorithms.

The data streams are buffered on local hard disks, with enough capacity to store the data for several days. This local buffer has multiple functions: 
\begin{itemize}
    \item It averages out the changes in data rate from luminosity changes so that the upstream link only need to provide average, not peak, bandwidth.
    \item It allows stand-alone operation for a limited time when the data transport out of the counting house is not available or runs at reduced capacity.
    \item It allows for near-online monitoring and replay of recent data for quality control, especially for those quantities which depend on on-going calibrations.        
\end{itemize}

The data are then pushed downstream to on-site or federated storage as part of the overall EIC project.

\subsection{Online monitoring}\label{subsec:online_mon}
Online monitoring is divided into a fast path with bound latency and a slow path. The fast path provides low-latency feedback for accelerator steering and equipment protection. The data for this path are generated early in the DAQ chain, either on the FELIX host CPUs or on the FELIX card themselves. Necessarily, they are limited in scope to counting, summing or similar type of information. The slow path provides higher-level information for quality control. Here, it's possible to reconstruct and analyse full events on-the-fly, by copying pre-selected time segments from the data stream to a dedicated server that performs full event reconstruction. Note here that such a monitoring system does not require the guaranteed reconstruction of all data, just of a suitable subset. That subset can either be selected unbiased by selecting periodic time segments, or biased by selecting time-segments tagged by data filters in the main data stream.  Similar monitoring can be performed on data on the local buffer for those quantities which require calibration data or two-pass analysis.

For both types of data a system will be needed to evaluate the monitoring data and inform the experiment operators of potential issues. This will largely include the creation of histograms which may be monitored either graphically or by some automated means. The prevalence of AI will certainly play a large role in that it will be able to evaluate a wider variation of monitoring data and at a much higher rate than could be expected of humans. Such systems are already deployed and under development~\cite{Hydra2021}. 

\subsection{Risk mitigation}
We expect that during initial commissioning noise rates will be significantly higher than during established operation, as accelerator and detector parameters will not yet be tuned optimally. Such high noise rates might overwhelm processing and upstream write capability. To allow progress in this initial phase, the DAQ system will accept as input a bounded-latency signal on the FELIX card or host CPU level to suppress uninteresting time segments.

Such a system also allows us to simply incorporate a dedicated collision detector for rate reduction: only time segments which are flagged to have a collision are kept, others are dropped early in the processing chain. 

This bounded-latency system could either be realized as a classic hardware filtering signal, or via software messages sent to the FELIX hosts with a clear advantage with regard to flexibility and ease of implementation, but with possibly a larger latency. The optimal implementation depends on the capabilities of the future FELIX successor and bandwidth availability on the FELIX host servers.

\subsection{AI-based data selection}

Traditionally, online data selection is performed using fixed topological cuts such as an energy deposit threshold in a calorimeter or the minimum transverse momentum of a track. Online data selection can benefit from the introduction of ML techniques typically used in offline data selection. It is advantageous to introduce these techniques at this stage to keep events that would be rejected by conventional methods and thereby increase the overall physics output of a detector. This approach can also simultaneously reject events that have a level of noise that would render a physics analysis of this event impractical.

The consortium is constructing a system where these selection techniques can be realised directly in hardware and will initially be deployed at sPHENIX which will double as a test system for an EIC detector. This project involves two symbiotic AI systems; one to identify collisions containing heavy flavor decays through their unique topology and another to determine the beamspot for the detector during operation. The latter system then feeds back to the data selection system to improve the physics efficiency.


In the system, the accept decisions would be handled by a separate FELIX system which aggregates information from several subdetectors but it should be noted that the choice of technology could evolve along with general hardware developments in the next decade. The FELIX has 48 bi-directional links allowing for data from multiple systems to be passed to a single FELIX board. The machine learning algorithm will then be loaded onto the FPGA which will be capable of basic tracking (using tracklets from the vertex, sagitta, and forward silicon tracking detectors) to make decisions that can be fed back to the initial DAQ or global data selection system to signal processing should continue. This should be achieved within 6$\,\mu$s\footnote{This requirement is determined by the shaping time of sPHENIX's vertex detector and could differ for ECCE}. 

%

Studies are ongoing, comparing the outputs of algorithms trained using Convolution Neural Networks (CNNs) and Graph Neural Networks (GNN). For many applications, GNNs often represent a more robust ML algorithm than CNN due to using additional information from the edges of the graph as well as the node information used by a traditional CNN. This makes them more applicable to sparse data such as tracking and data selection where the overall occupancy of the detector system is low.

It is important to note that the beam conditions can change with time, this can be both during a run and on a longer scale. Thus it is imperative that, as well as developing a selection algorithm, we develop a feedback system that is capable of monitoring the beam conditions in real time and adapting the input parameters to this. For example, the position of the collision point directly impacts the measurement of the track displacement. If this moves, it will alter the signal and background efficiencies in opposite directions. This monitoring will be achieved using GPUs which will feedback to the selection system. GPUs typically perform well using CNNs which motivates the study of various machine learning methods to find the optimal set of algorithms to use for each stage of the selection and monitoring. Other important features that can appear in a detector, impacting this system and hence must be monitored are the appearance of noisy channels (or pixels) and displacement of parts of the detector, such as through thermal expansion or vibrations. It has been demonstrated that algorithms can run fast enough on GPUs to achieve this form of monitoring~\cite{gpu2021}.


%

\subsection{Expected data rates and reduction steps}

Since connections between detector and FEPs might not be zero-suppressed digital or even analog data, it makes no sense to specify a data rate at the detector-to-hall border. Instead, the following section describes the expected data rate on the Fiber/FELIX level and downstream from there.

At nominal luminosity, we expect that true signals and beam-gas interactions produce a total rate of $\mathcal{O}$(100 Gbps) of zero-suppressed data at the FELIX card level. However, detector noise and additional backgrounds, especially during early operation, can completely dominate this rate. We assume therefore a total rate of $\mathcal{O}$(10 Tbps) bandwidth on the fiber level. Next-gen FELIX cards will have 25 Gbps receivers; for headroom for burst rates, non-ideal allocation of detector channels to fibers \emph{etc.}, we assume 12.5 Gbps as the average rate per fiber, and as a consequence, about 800 fibers.

A current generation FELIX card has ports that support 48 fibers. Assuming the same number of ports, this leads to O(20) next generation FELIX cards. Non-optimal distribution of FEE bandwidth across the fibers, the uncertainty in achievable reduction rate in the FELIX and achievable out-bound bandwidth may grow the number of required FELIX cards up to 3 fold for a total of 60. Each card would then receive 600 Gbps. We assume that at the time of procurement for EIC, cards based on at least PCIe Gen5 are available, providing 500 Gbps to the host server, requiring a modest reduction of the data rate in FELIX card itself, for example via cross-channel noise reduction. In combination with the host CPU, we expect a total reduction by a factor of 5 to $\mathcal{O}$(2) Tbps total, 100 Gbps per server. We note that a typical server with 128GB of memory can buffer the full stream for about 2 seconds, ample time for region-of-interest/time-slice-of-interest communication between the FELIX hosts, making higher reduction factors comparatively easy to achieve. The data can then be streamed out via a dual 100 Gbps link to the second layer in the compute farm.

In the compute farm, the data is further analyzed and filtered. We expect that with inter-detector noise suppression and high-level data selection the required effective bandwidth to long-term storage can be reduced to $\mathcal{O}$(100) Gbps.


\section {Offline}
\label{sec:offline}


Offline software encompasses many aspects of any experiment. This includes a number of systems, each of which requires either new development or implementation of existing systems using dedicated experts(s). These include:

\begin{itemize}
    \item Calibration system and database
    \item Reconstruction framework
    \item Reconstruction algorithms
    \item Simulation
    \item Offline Monitoring system
    \item Reconstruction workflow (HPC/HTC job management)
\end{itemize}

We intend to develop an offline computing model that aims for ``real time analysis'' that performs a single reconstruction pass on the data, producing reduced DSTs that are available for physics analysis on the time scale of a few weeks. In this description, the single reconstruction pass includes any relevant calibrations that are determined from specific calibration data sets. 

In the following sections we describe some of the above systems that will constitute larger efforts in terms of person-hours. It should be noted that at this time certain technology choices seem likely (e.g. GEANT4~\cite{ALLISON2016186}). However, others such as the choice of database systems, file formats, and software frameworks are purposefully left unspecified at this point in time. It is a primary goal post proposal period to define requirements and resource needs for these tasks.

\subsection{Reconstruction}\label{subsec:reconstruction}

In the past several decades, many reconstruction frameworks have been developed by different experiments within both HEP and NP. Several features stand out as common to all of these, which the ECCE software framework must utilize. The most important of these are modularity and user friendliness, as any large HEP/NP collaboration will necessarily comprise many hundreds of scientists with varying levels of software expertise. Therefore, these, and other generic features of excellent software, will be essential. It will additionally be imperative to recognize that software technologies change rapidly, and the ability for the software ecosystem to pivot with ease will be essential. As an example, while \texttt{git} is the de facto modern standard for code versioning and storage, it is impossible to say what versioning technologies will exist ten or more years from now when the EIC will be taking data. ECCE has not committed itself to a particular software ecosystem yet; however, these decisions will need to be made soon in preparation for development of a TDR.

One of the requirements of reconstruction software is reproducibility. ECCE will archive several daily builds that will provide users with the latest snapshot of the software; additionally, weekly builds that persist for longer periods of time will allow tracking of code evolution. In conjunction, special tagged production builds will be archived for large centrally produced data samples, such as those that were produced in preparation for the ECCE proposal. Currently the tagged releases are performed based only on time (e.g. weekly builds). Future software versions will consider implementing modern versioning practices such as semantic versioning~\cite{semantic}. In addition to archived builds, continuous integration is another tool ensuring reproducibility. ECCE has deployed continuous integration in certain repositories; however, automated tools enabled by services such as Jenkins or GitLab Runners will be deployed utilizing code checking tools and benchmark physics analyses. 

Making software user friendly requires that it is distributed in a convenient way. Currently the ECCE framework is distributed with \texttt{cvmfs}, a package managing software developed at CERN~\cite{cernvm}, while the software environment is containerized and deployed with Singularity~\cite{singularity}. Any software system that ECCE decides on will necessitate these tools for distribution, to ensure that all users can easily access the software and that a reproducible environment is available when deploying offline analysis and simulation in a federated computing architecture.

The role of hybrid architectures should also be considered in the ECCE reconstruction framework. Specifically, the use of GPU architectures will be important both for integrating machine learning into reconstruction workflows as well as generically taking advantage of the significant computational speed improvements that GPUs can provide, for example in charged particle reconstruction~\cite{Ai:2021kzk}. This integration has the added benefit of potentially utilizing the various leadership computing facilities that are available at national laboratories around the country, for more see Section~\ref{sec:offsite}.

Based on the experience of other experiments, reconstruction software should also take advantage of common software projects that are deployed across the world. For example, the A Common Tracking Software (ACTS) package, initially designed for use at the HL-LHC, has been implemented into the sPHENIX track reconstruction framework~\cite{Osborn:2021zlr}. Several collider-physics-based open source projects exist within the broader HEP community and have recently grown in their user base, examples include ACTS~\cite{Ai:2021ghi}, Rucio~\cite{Barisits:2019fyl}, PanDA~\cite{pandaDocs}, Fun4All~\cite{fun4allGithub}, JANA~\cite{jana2_chep19}, Gaudi~\cite{Gaudi}, and others. These should be evaluated for use within the ECCE software stack in 2022 as a part of the decision making process for the future of the offline software framework.

\subsection{Calibration}










Timely delivery of high-quality calibrations are one of the main challenges for EIC experiments, in particular given that many EIC measurements will be systematic uncertainty driven \cite{YellowReport}. ECCE adopts a fixed-latency production model, which requires the final calibration within 2-3 weeks of data taking. This leads to the design of a semi-automatic calibration workflow with minimal human intervention. There is already ongoing work to improve calibration workflows by integrating AI~\cite{Jeske:2022nws}.

Similar to the architecture of the sPHENIX computing model, we envision an offline computing center will provide a large incoming data buffer (e.g. 20PB as in the sPHENIX) that allows raw data to be used for reconstruction within 2-3 weeks of data taking, during which calibration will take place. The calibration tasks and time scale are dependent on the detector subsystems.

\subsubsection{Track Reconstruction}

For hits in the tracking detectors, the amplitude and time offset of each tracker channel will be aligned to a uniform response using an ensemble of collisions. We expect the initial calibration to be delivered within two weeks of data taking with frequent checks and updates when needed. 

\subsubsection{Particle Identification}

 Particle ID requires gain and time calibration. The single-photon and multi-photon per pixel hit from signal hit and noise will be used to set the gain. We expect a rapid turnaround for calibration and monitoring of the gain of approximately one day. The time offset calibration will be initially set by calibration-specific pulse lasers, which are applied before physics data taking. The final alignment requires events with a high multiplicity of tracks and aligning their projected collision time by adjusting timing shifts for each sensor, which will be part of the 4D alignment to be discussed at the end of this section.

 \subsubsection{Calorimetry}
 
 Calorimetric calibration focuses on the gain calibration. The first order of calorimetric energy scale calibration will be performed during production stage using the calorimeter blocks and SiPM QA database, e.g. light yield and gain measurements. The first iteration for the calorimetric energy scale will be based on cosmic data during the construction phase (e.g. sector testing) and pre-collision cosmic runs. This is expected to be completed before physics data taking. The second iteration of tower-by-tower energy scale variation calibration will be matching the energy slope of the calorimeter tower energy spectrum for the same eta slice. This is expected to be completed within one week of physics data taking. The final energy scale iteration will utilize real collision data in several channels. The first is using scattered electrons, $\pi^{\circ}\rightarrow\gamma\gamma$ and $\eta\rightarrow\gamma\gamma$ decays to set the energy scale for the EMCal. The second is using isolated hadronic shower to calibrate the e/h in EMCal and hadronic energy scale in the HCal. The third is using semi-inclusive deep-inelastic scattering single high momentum jet production to set the calorimetric jet energy scale. This is expected to take one week of data ($O(100)$ Billion events) and one week of calibration. During steady-state running, the tower-by-tower gain drift will be monitored and calibrated using LED flashes and SiPM temperature monitoring, which can be calibrated in about one hour.

\subsubsection{Alignment}

To fully align the entire detector, each subdetector will be surveyed before and after installation which provides the starting point of the alignment. The first iteration of alignment will use field-off data and cosmic data to adjust major pieces of the detector component to the final installed location. The time latency needed for this task is limited by the availability of such specialized data, but we expect this step is completed before the physics quality data taken at ECCE. The second iteration requires field-on physics quality collision data to provide the final high precision adjustment for the sensor locations and time offset (for TOF) to a small fraction of the resolution. The first period of magnetic-field-on collision data will be used for this alignment. Generic purpose global alignment package such as Millepede II will be used. Other packages, such as alignment software available in ACTS~\cite{Ai:2021ghi}, can be considered as well. It is expected to take two weeks to complete the iteration of alignment and checks. Steady-state updates: the vertex tracker requires O(1)~$\mu$m alignment precision, which could change over long periods. Therefore, during steady-state running, we expect alignment to be checked every few days and possibly updated every week, depending on the final mechanical stability. We expect steady-state alignment updates can be achieved within one hour (e.g. the LHCb vertex tracker is aligned in about 7 minutes~\cite{BORGHI2017560}).

\subsubsection{Calibration database}

For the ECCE streaming DAQ, we expect the calibration record to be time-stamped with a 64-bit beam crossing counter with the start and end time corresponding to the interval of validity. The validity window length will be detector and calibration dependent, but we expect they align with the luminosity block of ECCE streaming data that is $\mathcal{O}(1)$ second, at each of the electron ring bunch refills. 

The size of the calibration data is much smaller than the raw data but still sizable. For the highest channel count in the silicon vertex tracker (O(1)B channels), we do not expect to have frequent calibration as it presents boolean (hit/no-hit) pixel data. As a conservative estimate, consider 300 thousand of the calorimeter, tracker, and PID channels that require a frequent (1 per minute) update of a relative gain and time shift, each represented by a 4-byte float. This gives an overall calibration dataset size of $O(1)$~TB per run year (8B*(300e3 chan)*60(minute)*24(hour)*7(day)*20(cryo week)). 

The calibration data will be indexed in a relational database. The actual calibration data files can exist in a distributed file server (e.g.\ S3 or XROOTD~\cite{xROOTD}) or in a separated database table, depending on the size per entry and frequency of calibration updates. The separation of index database and calibration data payload allows for efficient database implementation management that is capable of accommodating a possible large size of the calibration data. This approach is being deployed in the sPHENIX fixed latency calibration and reconstruction.

At the start of a production job, the job manager will pull the calibration data relevant for the job into the local disk buffer according to the index table. Then the database is disconnected and job processing starts to efficiently utilize the connection limit to the database. This also allows flexibility to pre-assemble the calibration file package to be sent with raw data to remote computing centers that are otherwise disconnected from the ECCE database and file servers.

\subsection{Simulation}






	All modern high energy physics experiments require highly detailed detector simulations, both in the design and operational phases. The volume of simulations required depends on their specific uses, from small scale simulations with hundreds of events to study new sub-detector systems to large scale simulations with over 100 million events to understand physics capabilities. With this in mind, the ECCE philosophy towards simulations is user-friendliness, modularity, and no distinction between large and small scale simulations to avoid disparities. 
	
	The current framework is capable of performing simulations from the generator stage right up to final physics analysis, with intermediate stages for detector simulation, responses, and track, PID and calorimeter reconstruction. This benefits users by allowing them to run a full analysis chain in one step if desired and large scale productions by breaking simulations down into a series of stages. The latter approach improves throughput and reproducability as the same generator-level simulation can be run over different detector configurations or more physics objects can be added later in time when for example the simulation and reconstruction are run separately. It is expected that any new framework used will try to retain as much of these features as possible.
	
	The framework has several inbuilt event generators (a particle gun, \textsc{Pythia6} \cite{Sjostrand:2000wi}, \textsc{Pythia8} \cite{Sjostrand:2007gs} and \textsc{Sartre}~\cite{Toll:2013gda}) and can also read in pre-generated events either via the EIC-smear program~\cite{eicsmear} or a file in \textsc{HepMC2} format~\cite{Dobbs:2001ck}. The framework is also capable of reading in any previously produced DST, assuming the material hits were saved. If any generated particle has not been decayed by the input generator and is required to decay in the detector volume, this is handled by the built-in \textsc{Pythia6} decayer. 
	
	The detector simulation will likely be handled by \textsc{Geant4}~\cite{Agostinelli:2002hh}. The components of the detector can either be rendered in what is called "fast" or "full" simulation. Fast simulation is useful for producing passive volumes such as support or service structures, or testing ideas quickly. Full simulation will involve complete physics responses and digitization, including but not limited to Cherenkov photon production and electron cascades. 
	
	Efforts are on-going to improve both our simulations, through work conducted with AI-assisted detector designs~\cite{ecce-note-comp-2021-03}, and the infrastructure needed to produce large simulations on short time scales such as with distributed computing and GPU implementations of \textsc{Geant4}. Each individually simulated subsystem is bundled with the software stack which means it is also saved weekly and with every production build. This allows any simulation to be reproduced at a future date, if necessary, and this ability should be maintained throughout the experiment lifetime and beyond. 
	
    The AI-assisted design optimization of the ECCE inner tracker \cite{ecce-note-comp-2021-03} has been based on evolutionary algorithms; during the detector proposal multiple optimization pipelines have been run each with a population size of 100, representing different detector design configurations. At each iteration, AI updates the population. The total computing budget for an individual pipeline amounted to approximately 10k CPU-core hours. 
    Activities are planned to continue the detector optimization: new optimization pipelines can deal with larger parameter space to include a system of sub-detectors like in the case of the whole ECCE tracker  \cite{ecce-note-comp-2021-03}; we also plan to optimize other sub-detectors like, \textit{e.g.}, the d-RICH, leveraging on the expertise internal to the ECCE collaboration regarding specifically the design of the dRICH with AI-based techniques \cite{cisbani2020ai}. 
    Larger populations may need to be simulated to cope with the increased complexity in order to improve the accuracy of the approximated Pareto front. Different AI-based strategies will be compared. 
    We anticipate roughly 1M CPU-core hours per year for these AI based studies.

	\begin{figure}[ht]
		\begin{center}
			\includegraphics[width=0.49\textwidth]{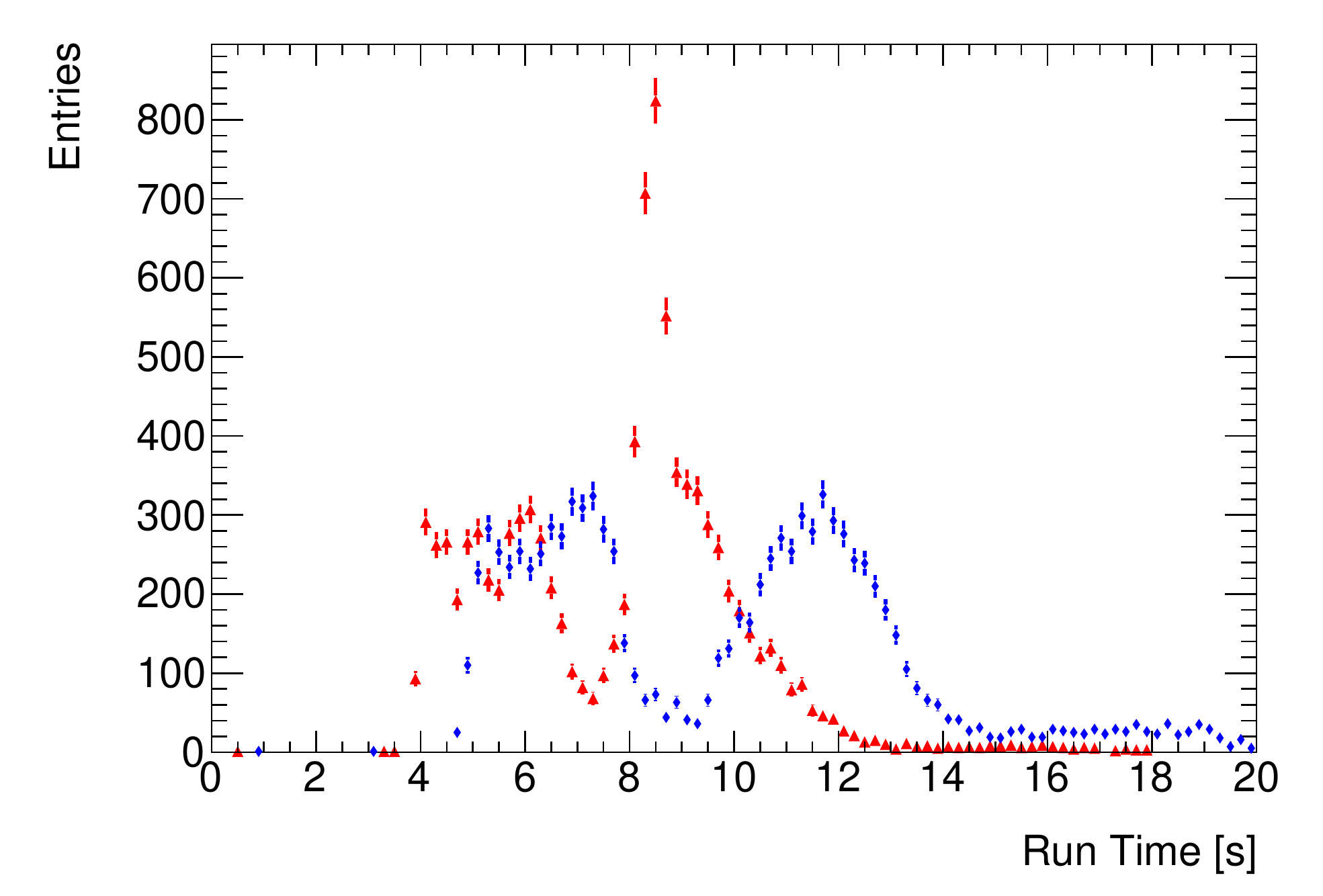}
			\includegraphics[width=0.49\textwidth]{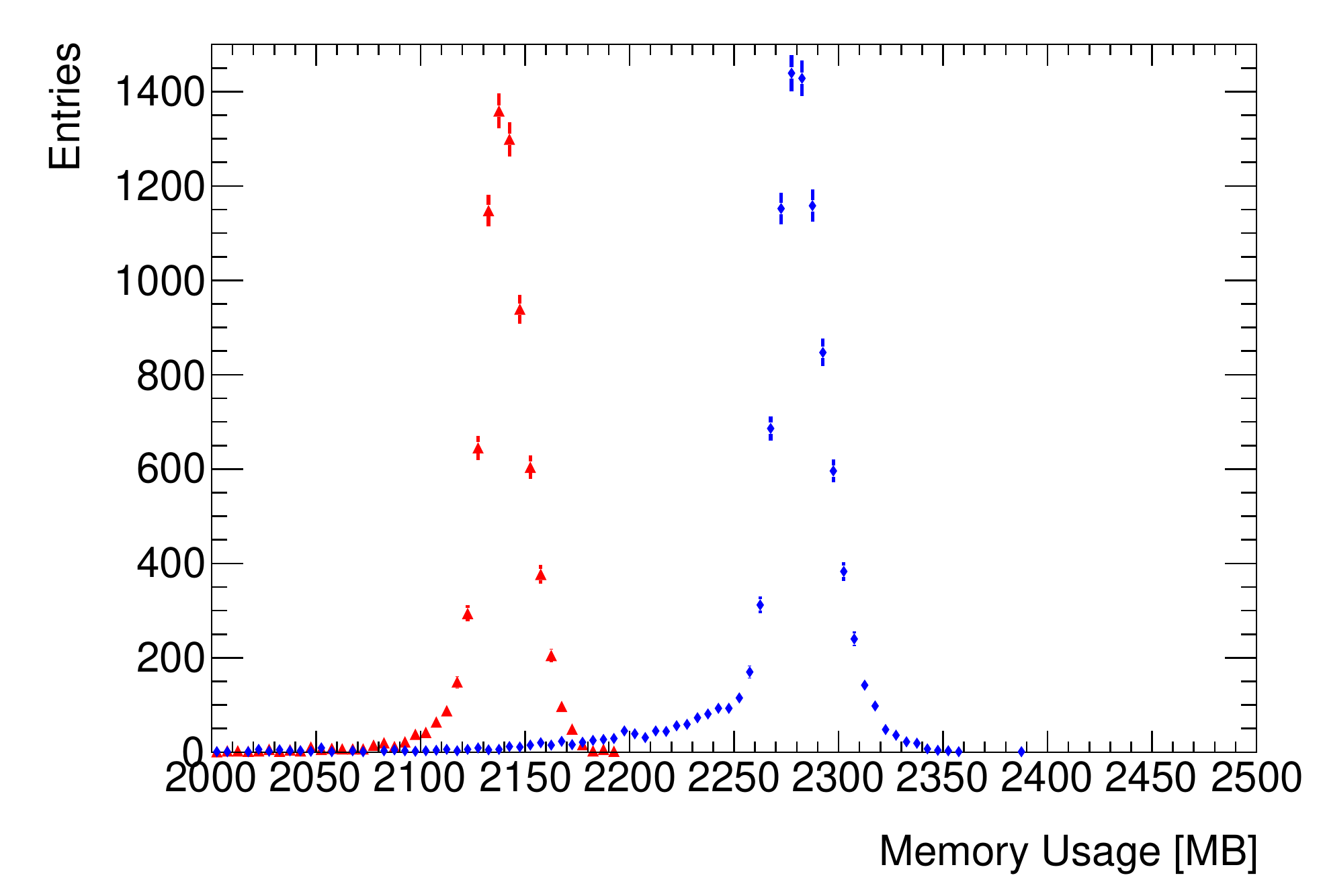}
		\end{center}
		\caption{\small The per-event run time (top) and per-job memory usage (bottom) for two different productions. $ep$ collisions with a 10~GeV electron beam and 100~GeV proton beam using an internal \textsc{Pythia8} generator are shown with red triangles while $ep$ collisions with a 18~GeV electron beam and 275~GeV proton beam using an external \textsc{Pythia6} generator are shown with  blue diamonds.  As each entry in the run time is the average time to produce 2000 events the multiple peaks for each production is due to different hardware used to process the jobs on the batch farm.}\label{fig:sim_jobs}
	\end{figure}
	
	The ECCE consortium conducted two large scale production campaigns in 2021; the first campaign consisted of over 120M events while the second campaign consisted of over 600M events. The campaigns were distributed over 3 distinct production sites; SDCC at Brookhaven National Laboratory, the SciComp at Thomas Jefferson National Accelerator Facility and MIT Bates Research and Engineering Center. The production sites used a common top-level program which is able to communicate with site-specific lower level programs. With this and the common simulation framework, production tasks can be assigned to any site and down-time at one site can be recovered with a different site. As the only difference between the sites is in the batch systems, each production site is capable of creating output files and directories in identical formats and hence the production location is transparent to end users. Finally, the simulation seeds are uniquely defined by the input options (input file name, number of events to generate and starting event number of the input file) so any site can precisely reproduce any file from the other sites which aids in both debugging and general event production. As well as the large scale production at these three sites, another large production of almost 50M events was generated using computing resources at Oak Ridge National Laboratory for calorimetry development. This simulation used a different production mechanism from that discussed previously, demonstrating the flexibility and advantage of using this modular system.

	The second simulation campaign featured a far more mature detector design with full PID, calorimeter responses, optimal detector placements and support structures. Thus, this campaign is useful for bench-marking the simulation memory usage and processing times. By comparing several large productions, it was found that the average time to produce a single $ep$ event with a 10~GeV electron beam and 100~GeV proton beam using the in-built \textsc{Pythia8} generator is 7.8\,s with a standard deviation of 2.2\,s\footnote{These events involved generator level production, detector simulation, digitization, reconstruction (track, PID and calorimeter) and physics analysis output}. This value was obtained by studying approximately 20 million events, grouped into production jobs of 2000 events each and taking the average run time of each job. Thus, this number also includes the start up of the framework but this impact should be minimized by using the average of 2000 events. Similarly, the average memory usage of a job was found to be stable regardless of the number of events that were produced in each job. A small variation in run time is seen with respect to the collision energies which is expected; when the beam energies increase, the event multiplicity increases and hence there are more objects to simulate and reconstruct. This can be seen by simulating $ep$ collisions with a 18~GeV electron beam and 275~GeV proton beam using an external \textsc{Pythia6} generator and the internal EIC-smear reader which found an average event production time of 9.7\,s with a standard deviation of 3.3\,s. This is also reflected in the simulation memory usage where the collisions with a 10~GeV electron beam and 100~GeV proton beam had an average memory usage of 2138\,MB with a standard deviation of 16\,MB while a 18~GeV electron beam and 275~GeV proton beam had an average memory usage of 2275\,MB with a standard deviation of 32\,MB. It is also expected that the overall memory footprint will be reduced through code optimisation and new hardware. For example, it has already been demonstrated in sPHENIX (which shares the same framework) that the mean memory usage for $pp$ simulations can be reduced from 4\,GB to 1.7\,GB by selective loading of simulated materials. Currently, ECCE simulations load every material described in \textsc{Geant4} into memory. The distributions for event run time and memory use are given in Figure \ref{fig:sim_jobs}.

The campaigns performed in 2021 can be used to estimate the simulation requirements for the forthcoming years. These estimates are given in Tables~\ref{tab:sim_predictions} and \ref{tab:sim_predictions_data_taking} for the R\&D and data taking periods respectively. The first table assumes that a large production will occur in 2022 based on reviewers suggestions which will steadily decrease as detector R$\&$D progresses for several years before increasing significantly in the years leading to data taking as the collaboration performs as realistic simulations as possible to exercise the reconstruction, calibration and alignment software.The simulation requirements for the data-taking period assume that the collaboration will need $O(10)$ times the amount of simulated data for $O(10)$\% of the streaming recorded minimum bias cross section in the real data for each running year that is most relevant for the core physics program at ECCE. The number of expected real events recorded are listed in Table~\ref{tab:cpu_summary}, which is comparable to the computing need in the offline reconstruction as discussed in Section~\ref{sec:resources}.

\begin{table*}[ht]
	\centering
	\begin{tabular}{c|c|c|c}
		\hline
		Year & Number of Events [$\times 10^{6}$] & Storage [TB] & CPU-core hours [Mcore-hrs] \\
		\hline
		\hline
		2022 & 200 & 50 & 45 \\
		2023 - 2024 & 100 & 25 & 22.5 \\
		2025 - 2028 & 50 & 12.5 & 11 \\
		2029 - 2030 & 500 & 125 & 110 \\
		\hline
		\hline
		\textbf{Total} & 1600 & 400 & 354 \\
		\hline
	\end{tabular}
	\caption[]{Estimated simulation requirements for the years 2022 - 2030. The estimates are based on the observed performance in 2021, only include large scale productions and hence do not include any productions for AI-assisted detector design. The numbers assume that a large scale campaign will take place in 2022, based on feedback from the proposal. The productions will then decrease as focus moves into hardware development before increasing significantly in the years before initial data taking as "Mock Data Challenges" are pursued to test the reconstruction, calibration and alignment software.}
	\label{tab:sim_predictions}
\end{table*}

\begin{table*}[ht]
	\centering
	\begin{tabular}{c|c|c|c}
		\hline
		Year & Number of Events [$\times 10^{9}$] & Storage [PB] & CPU-core hours [Mcore-hrs] \\
		\hline
		\hline
		year-1 & 120 & 30 & 11000 \\
		year-2 & 600 & 150 & 55000 \\
		year-3 & 5400 & 1300 & 490000 \\
		\hline
	\end{tabular}
	\caption[]{Estimated simulation requirements during operational years. The storage and CPU time estimates are based on the observed performance in 2021 while the number of events assume we will need 
	$\mathcal{O}(10)$ times the amount of simulated data for $\mathcal{O}(10)$\% of the streaming recorded minimum bias cross section in the real data for each running year that is most relevant for the core physics program at ECCE
	}
	\label{tab:sim_predictions_data_taking}
\end{table*}

\section {Offsite Processing}
\label{sec:offsite}

The EIC will be a large international project with many researchers and stakeholders spread throughout the globe. While the accelerator and detectors are necessarily placed at a single locale, the computing need not be and can better reflect the geographic diversity of the collaborations involved in EIC research. In the modern age, high speed network connectivity has become very robust. In the time frame of the EIC ($\sim 2030$) we can expect even more reliable and even faster networks. This will make transporting data on the scale the EIC is expected to produce fairly routine. To give specific numbers, BNL has a 400Gbps connection to the ESnet backbone in 2021. ECCE is currently estimated to produce 100Gbps of raw data once it is in full production sometime around 2030. Network bandwidths have shown steady growth of about 50\% per year over the past few decades\cite{nielsensLaw2019}. Thus, over the next 8 years we can expect existing bandwidths to grow by roughly a factor of 25. Even a conservative estimate that the BNL external connection bandwidth grows by only a factor of 10 means the entire ECCE raw data volume can be streamed out using only a few percent of the total available bandwidth.

This section will briefly describe how a federated computing model for the EIC might look, how it will be used to process the raw data, and how it will also be used to process the large amounts of simulation needed for the program.

\subsection{Federated computing}

EIC data processing will employ a federated computing model where multiple facilities will be used. A similar strategy has been successfully deployed by the LHC in the form of the Worldwide LHC Computing Grid (WLCG) or simply \emph{the Grid}\cite{SHIERS2007219}. The benefits of distributing the computing across multiple sites include:

\begin{itemize}
    \item Each site only needs to handle a fraction of data
    \item EIC computing becomes a smaller fraction of each compute farm
    \item One site having diminished capacity temporarily can easily be absorbed by others without reconfiguration
\end{itemize}

Generally speaking, diversifying helps to mitigate certain risks. 

The WLCG model of the LHC is based on multiple \emph{Tiers}, structured in a pyramid type fashion. The topmost tier (\emph{Tier 0})represents the LHC/CERN where the data is produced and all data is stored. It also supplies around 20\% of the total computing resource and does the initial reconstruction of the data before distributing it to the Tier 1 sites\cite{WLCG_Tiers_website}. The Tier 1 sites perform large scale reprocessing of the data and distribution to the Tier 2 sites. The Tier 2 sites do more specialized analysis and simulations while the Tier 3 sites are end users. Each tier in the system only communicates with tiers directly above or below it in the hierarchy.

\begin{figure}[hbt!]
 \begin{center}
   \raisebox{0.5mm}{\includegraphics[clip, width=0.99\linewidth]{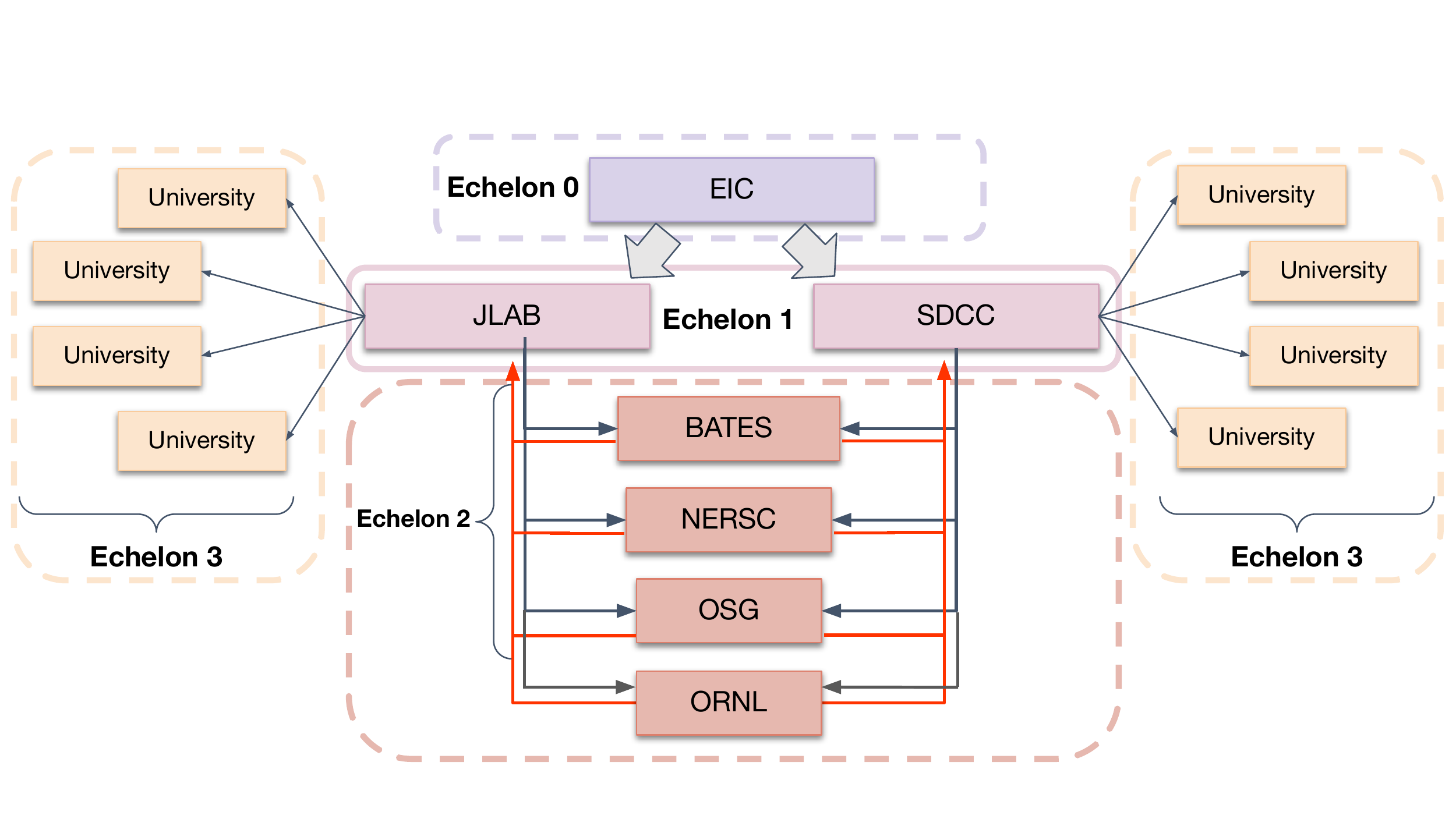}}
  \caption[Federated Computing Butterfly model.]{\label{fig:federated_offsite_butterfly} Butterfly model of federated offsite computing. In this model, nearly all storage is contained in echelon 1 while large portions of the raw data processing is delegated to multiple HTC/HPC facilities. The named facilities in this graphic are merely examples and do not represent commitments or final plans.}
 \end{center}
\end{figure}

For the EIC one may consider an alternative model in which the data producer (the experiments of the EIC) and the BNL compute facility (e.g. SDCC) are independent. This allows computing at BNL to become part of a pool of facilities that handle the computing as a federated resource. Figure \ref{fig:federated_offsite_butterfly} illustrates such a model referred to here as a ``\emph{Butterfly}'' model due to the rough shape of the figure. In this model, both compute and storage are distributed with the storage being focused in the Echelon 1 sites. This means access to the data by the end users will be done by connecting Echelon 3 sites directly to Echelon 1 sites. The Echelon 1 sites will themselves provide significant compute capability, but may also farm out large campaigns to Echelon 2 sites. In the simulation campaigns performed for the ECCE proposal, the model shown in Fig.~\ref{fig:federated_offsite_butterfly} was successfully implemented for simulated data production.

\subsection{Raw data compute}
\begin{figure}[hbt!]
 \begin{center}
   \raisebox{0.5mm}{\includegraphics[clip, width=0.99\linewidth]{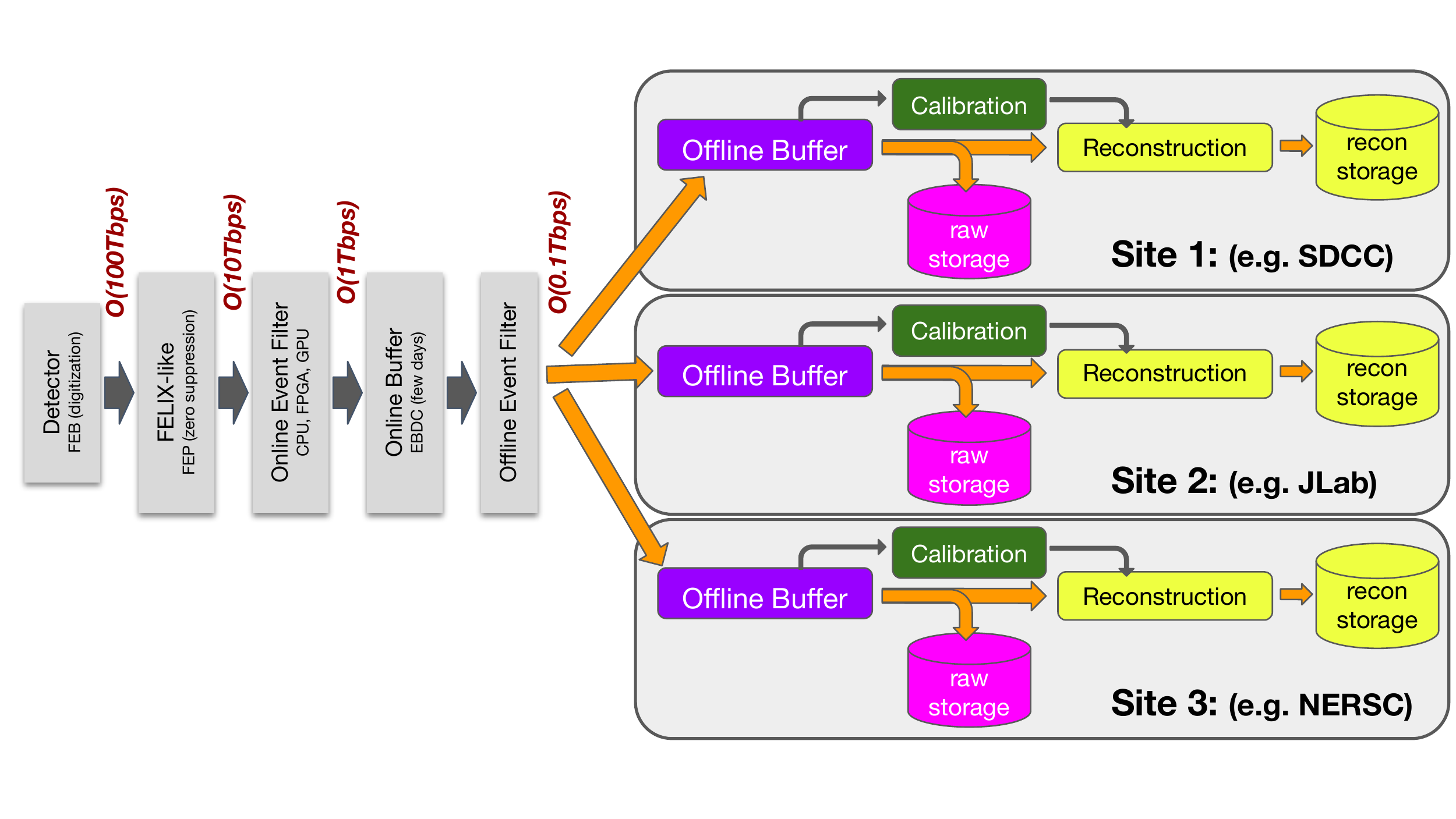}}
  \caption[Example of federated processing of data in near real-time using multiple sites.]{\label{fig:federated_offsite_example} Data flow from detector to reconstructed object files (left to right). This diagram illustrates how raw data may be distributed to multiple sites in near-real time. On the left side of the plot, multiple filter and buffering stages are used to reduce the data rate. On the right, the data is distributed to multiple facilities.
  Each facility would store a portion of the raw data. It would also need to keep the data live (e.g. on disk) long enough for it to be calibrated and then processed by the reconstruction software.}
 \end{center}
\end{figure}

Processing of EIC data will occur over multiple sites which will include HTC facilities at both BNL and JLab and possibly others. The plan calls for processing the raw data into reconstructed objects such as tracks, jets, and calorimeter clusters within 2-3 weeks of acquisition. The bulk of the few week latency will be due to the time it takes to calibrate the data so that reconstruction may occur. Figure \ref{fig:federated_offsite_example} illustrates how such a scheme could work. The raw data read from the streaming DAQ system will need to be reduced over multiple filtering and compression stages to a rate that is reasonable to transport offsite from BNL using its external network connection. Table \ref{tab:reduction_factors} lists the stages and with in-going and out-going rates and their respective reduction factors. Potential technologies that could be applied at each stage are also listed.

The DOE lab systems are connected via the ESNet unclassified network for scientific research~\cite{ESNet}. As of 2021, BNL has a 400Gbps connection to ESNet and JLab has dual 10Gbps connections. In 2022 or 2023, JLab is anticipated to increase its bandwidth to at least 100Gbps. When the EIC begins collecting data around 2030, one may expect a 1Tbps bandwidth between the two labs. This is an order of magnitude higher than the anticipated raw data rate after filtering from the ECCE detector. Thus, transfer of the entire raw data set offsite from BNL in 2030 seems reasonable.

\begin{table*}[ht]
    \centering
    \begin{tabular}{p{4cm}|c|c|c}
        \hline
        Stage                        & Input/Output    & Reduction Factor & Technology options \\
        \hline
        \hline
        Compute Interface (e.g. FELIX) & 100Tbps/10Tbps  & $\times 10^{-1}$ & FPGA \\
        \hline
        Online Event Filter          & 10Tbps/1Tbps    & $\times 10^{-1}$ & FPGA, (GPU), CPU\\
        \hline
        Online Buffer                & 1Tbps/0.5Tbps   & $\times 5x10^{-1}$  & $<disk>$ \\
        \hline
        Offline Event Filter         & 0.5Tbps/100Gbps & $\times 2x10^{-1}$  & FPGA, GPU, CPU \\
        \hline
        Reconstruction               & 100Gbps/10Gbps  & $\times 10^{-1}$ & (FPGA), GPU,CPU\\
        \hline
        \hline
        \textbf{Total}               & \textbf{100Tbps/10Gbps} & \textbf{$\times 10^{-4}$} & \\
        \hline
    \end{tabular}
    \caption{Data rates and reduction factors for proposed near real time data flow. Estimated data rate from ECCE detector is $\mathcal{O}(100Tbps)$. Raw storage will be $\mathcal{O}(100Gbps)$. Reconstructed object storage will be $\mathcal{O}(10Gbps)$. Parentheses indicate technologies that could be used, but seem less likely choices.}
    \label{tab:reduction_factors}
\end{table*}


\section {Resource Requirements Summary}
\label{sec:resources}

The EIC luminosity is projected to be between $10^{33}cm^{-2}s{-1}$ and $10^{34}cm^{-2}s{-1}$ (see sec. 2.10 of the Yellow Report~\cite{YellowReport}). Assume 30 weeks of operation per year and 60\% accelerator operation efficiency once it is in full production mode. In the first years, however, we may expect fewer weeks of running and lower luminosity. Table \ref{tab:integrated_luminosity_by_year} lists a possible scenario used for the purposes of estimation in this section. We assume 100Gbps data rate to storage for $10^{34}cm^{-2}s{-1}$ and 60\% operational efficiency of the facility. All other rates are derived by scaling this value by the luminosity and efficiency values indicated in the table.

\begin{table*}[htb]
    \centering
    \begin{tabular}{c|c|c|c}
        \hline
        \hline
         \textbf{New Storage}       & year-1                & year-2                  & year-3                \\
        \hline
         Luminosity              & $10^{33}cm^{-2}s^{-1}$ & $2\times10^{33}cm^{-2}s^{-1}$ & $10^{34}cm^{-2}s^{-1}$ \\
         \hline
         Weeks of Physics Running        & 10                    & 20                      & 30                    \\
         \hline
         Operational efficiency    & 40\%                  & 50\%                    & 60\%                  \\
         \hline
         Data Rate to Storage    & 6.7Gbps               & 16.7Gbps                & 100Gbps               \\
         \hline
         Raw Data Storage (no duplicates) & 4PB          & 20PB                    & 181PB                 \\
         \hline
         Recon Storage          & 0.4PB                  & 2PB                    & 18PB                   \\ 
         \hline
         Total Storage (no duplicates) & 7PB           & 35PB                   & 317PB                  \\
         \hline
   \end{tabular}
    \caption{Estimate of raw data tape storage needed for first 3 years of EIC running (ECCE only). Values are estimates assuming ramp up to full luminosity  by year 3. Numbers for the first two years are estimated for the purposes of this exercise and do not come from an external source. n.b. each value represents \emph{only} the needs for data produced in that year and \emph{not} a cumulative total.}
    \label{tab:integrated_luminosity_by_year}
\end{table*}

Temporary disk storage will be needed for raw data during the 3 week time span during which calibrations are derived and the raw data processed. In addition, disk storage will be needed for the reconstructed data that collaborators will be accessing for analysis. Table \ref{tab:disk_summary} gives estimates of the disk resources needed for the first 3 years of running. Note that the values in the table are cumulative and so represent the total amount of disk needed for each year which include reconstructed data from previous years.

\begin{table}[htb!]
    \centering
    \begin{tabular}{c|c|c|c}
        \hline
        \textbf{Total Disk} & year-1 & year-2 & year-3 \\
        \hline
        \hline
        Disk (temporary)  &  1.2PB & 3.0PB & 18.1PB \\
        \hline
        Disk (permanent)    & 0.4PB & 2.4PB &	20.6PB \\
        \hline
        \textbf{TOTAL}          & 1.6PB &	5.4PB &	38.7PB \\
        \hline
    \end{tabular}
    \caption{Estimate of disk storage needed for first 3 years of EIC running (ECCE only). The temporary disk is used to hold raw data for a 3 week period while calibrations are derived and reconstruction is done. The permanent disk is for holding the reconstructed data. This will be cumulative so collaborators will have access to recon data from all years.}
    \label{tab:disk_summary}
\end{table}

The CPU required for processing the data is very difficult to estimate with any accuracy better than the order of magnitude. Nonetheless, an attempt is made here to provide such an estimate. Table \ref{tab:cpu_summary} summarizes the important values. The 5.4s/ev comes from estimating an average of 3 hours for reconstruction of 2k events of ECCE data. The numbers for ECCE CPU mainly come from the simulation campaigns run for proposal development which include combined simulation and reconstruction. The times to process 2k events ranged from 2 to 9 hours depending on the collision type and the CPU type that the job was processed on. This corresponds to a range of roughly 4s/ev to 16s/ev. The reconstruction only part is considered to be half of the roughly 6 hour average time to simulate 2k events. By way of comparison, sPHENIX estimates 15s/ev for Au+Au scattering and 10.4s/ev for p+p scattering (see section 5.2 of \cite{sphenix_computing_plan_2019}). Thus, 5.4s/ev is assumed to be at least the right order of magnitude. The event size of 250kB is also a rough average based on the ECCE DST files for several configurations simulated in the major proposal campaigns. Note the DST event size is larger than the average raw data rate divided by the event rate, as the streaming readout raw data is more tightly packed in time-frames which also avoids duplication of information between the neighboring events. The event size is used, along with the numbers for the Raw Data Storage from table \ref{tab:integrated_luminosity_by_year}, to calculate the number of events produced in each year. The CPU needed for calibration is estimated to be roughly 5\% of that needed for full reconstruction. It is noted that the sPHENIX Computing Plan estimates this to be 25\%. The final line in table \ref{tab:cpu_summary} estimates the number of CPU cores needed to process the data for each year assuming it can be done over a 30 week period. This would mean in year-3 there would be enough CPU to keep up with the raw data production rate. In earlier years, this would not be needed as the production times are much shorter.

\begin{table*}[htb!]
    \centering
    \begin{tabular}{c|c|c|c}
        \hline
        CPU Compute & year-1 & year-2 & year-3 \\
        \hline
        \hline
        Recon process time/core	& 5.4s/ev	& 5.4s/ev	& 5.4s/ev \\
        \hline
        Streaming-unpacked event size	& 33kB	& 33kB & 33kB \\
        \hline
        Number of events produced &	121 billion	& 605 billion & 5,443 billion \\
        \hline
        CPU-core hours (recon-only, 1 pass)	& 181Mcore-hrs	& 907Mcore-hrs &	8165Mcore-hrs \\
        \hline
        CPU-core hours (calib-only) &	9Mcore-hrs &	45Mcore-hrs &	408Mcore-hrs \\	
        \hline
        2020-cores needed to process in 30 weeks	& 38k &	189k &	1701k \\
        \hline
    \end{tabular}
    \caption{Estimates of CPU needed for reconstruction of raw data. The number of seconds per event is highly dependent on the type of processor being used. Number of events comes from total raw data storage estimate in table \ref{tab:integrated_luminosity_by_year}. Calibration is assumed to be 5\% of reconstruction time.}
    \label{tab:cpu_summary}
\end{table*}


\section{Summary}
\label{sec:executive_summary}

The ECCE consortium plans to deploy a federated computing model for the EIC where multiple facilities are used. ECCE recognizes the need for a global EIC model and intends to fully participate in the design and implementation of such a system. A similar strategy has been successfully deployed by the LHC in the form of the Worldwide LHC Computing Grid (WLCG)~\cite{SHIERS2007219}. ECCE has developed and, during the EIC detector proposal period, deployed a tiered ``Butterfly'' model for EIC computing that was inspired by the WLCG model, but updated to better reflect the computing landscape anticipated for the EIC. In this model, the EIC detector supplies the data, but the SDCC at BNL is treated as one of a pool of sites used for long term storage and compute resources. Both BNL and JLab would be considered as \emph{Echelon 1} sites with the ability to add others as appropriate. Raw data would be distributed amongst multiple \emph{Echelon 2} sites for processing with the processed data being returned to Echelon 1. Researchers would directly access the processed data at the Echelon 1 sites.

We have adopted a fixed-latency offline computing model where both the final calibration and reconstruction of raw data occur within 2-3 weeks of acquisition. During this period, raw data will be buffered on disk at all of the Echelon 1 sites, along with permanent archival copies on tapes. Final calibration will be performed semi-automatically including accumulating sufficient data for tracker alignment and energy scale calibration of the calorimeters. Artificial intelligence and machine learning will be integrated throughout this model. After calibration, data processing  will be released to multiple sites including HTC facilities at both Echelon 1 and 2 sites. The EIC will also require large simulation samples to aid in understanding the detector response and physics and background processes being measured.
 We expect that the produced simulation sample will focus on 10\% of the EIC collision cross-section that is directly relevant for the signal and background of the core ECCE physics program. These physics processes will be simulated to $O(10)$~times the statistics in real data to constrain systematic uncertainty from the simulated sample to be much smaller than the data statistical uncertainty.

A summary of the anticipated resource requirements can be seen in table \ref{tab:computing-integrated_luminosity_by_year}.

\begin{table*}[ht]
    \centering
    \begin{tabular}{c|c|c|c}
        \hline
         \textbf{ECCE Runs}       & year-1                & year-2                  & year-3                \\
        \hline
         Luminosity              & $10^{33}\mathrm{cm}^{-2}\mathrm{s}^{-1}$ & $2\times 10^{33}\mathrm{cm}^{-2}\mathrm{s}^{-1}$ & $10^{34}\mathrm{cm}^{-2}\mathrm{s}^{-1}$ \\
         Weeks of Running        & 10                    & 20                      & 30                    \\
         Operational efficiency    & 40\%                  & 50\%                    & 60\%                  \\
        Disk (temporary)  &  1.2PB & 3.0PB & 18.1PB \\
        Disk (permanent)    & 0.4PB & 2.4PB &	20.6PB \\
         Data Rate to Storage    & 6.7Gbps               & 16.7Gbps                & 100Gbps               \\
         Raw Data Storage (no duplicates) & 4PB          & 20PB                    & 181PB                 \\
        \hline
        Recon process time/core	& 5.4s/ev	& 5.4s/ev	& 5.4s/ev \\
        Streaming-unpacked event size	& 33kB	& 33kB & 33kB \\
        Number of events produced &	121 billion	& 605 billion & 5,443 billion \\
         Recon Storage          & 0.4PB                  & 2PB                    & 18PB                   \\
        CPU-core hours (recon+calib)	& 191Mcore-hrs	& 953Mcore-hrs &	8,573Mcore-hrs \\
        2020-cores needed to process in 30 weeks	& 38k &	189k &	1,701k \\
        \hline
   \end{tabular}
    \caption{Estimate of raw data storage and compute needs for first 3 years of ECCE,  assuming ramp up to full luminosity by year 3. 
    }
    \label{tab:computing-integrated_luminosity_by_year}
\end{table*}

\section{Acknowledgements}
\label{acknowledgements}

We thank the EIC Silicon Consortium for cost estimate methodologies concerning silicon tracking systems, technical discussions, and comments.  We acknowledge the important prior work of projects eRD16, eRD18, and eRD25 concerning research and development of MAPS silicon tracking technologies.

We thank the EIC LGAD Consortium for technical discussions and acknowledge the prior work of project eRD112.

We acknowledge support from the Office of Nuclear Physics in the Office of Science in the Department of Energy, the National Science Foundation, and the Los Alamos National Laboratory Laboratory Directed Research and Development (LDRD) 20200022DR.





\bibliographystyle{elsarticle-num} 
\bibliography{references,refs-ecce}

\end{document}